\documentclass[11pt]{article}

\usepackage{acl}

\usepackage{times}
\usepackage{latexsym}
\usepackage{microtype}
\usepackage{tabularx}
\usepackage{makecell}
\usepackage{placeins}
\usepackage{caption}
\usepackage{needspace}
\usepackage[T1]{fontenc}

\usepackage[utf8]{inputenc}
\usepackage{algorithm}
\usepackage{algpseudocode}
\usepackage{microtype}
\usepackage{amsmath}
\usepackage{mdframed}
\usepackage[most]{tcolorbox}
\tcbuselibrary{breakable}
\usepackage{amssymb}

\definecolor{QEBlue}{HTML}{2F5DA8}
\definecolor{QELightBlue}{HTML}{F4F7FC}
\definecolor{QEGray}{HTML}{4B5563}
\definecolor{QEOrange}{HTML}{B45309}

\newcommand{\promptrole}[1]{{\sffamily\bfseries\color{QEBlue}#1}}

\lstdefinestyle{promptjson}{
    basicstyle=\ttfamily\footnotesize,
    columns=fullflexible,
    keepspaces=true,
    breaklines=true,
    frame=none,
    numbers=none,
    framerule=0.25pt,
    rulecolor=\color{QEGray},
    xleftmargin=0.5em,
    xrightmargin=0.5em,
    aboveskip=0.4em,
    belowskip=0.2em
}

\newtcolorbox{qualpromptbox}[1]{
    enhanced,
    colback=QELightBlue,
    colframe=QEBlue,
    coltitle=white,
    fonttitle=\bfseries\sffamily,
    title=#1,
    boxed title style={
        colback=QEBlue,
        colframe=QEBlue,
        arc=1.5mm,
        boxrule=0pt
    },
    attach boxed title to top left={
        xshift=3mm,
        yshift=-2mm
    },
    arc=2mm,
    boxrule=0.7pt,
    left=3.5mm,
    right=3.5mm,
    top=4mm,
    bottom=3mm,
    before upper={
        \small
        \setlength{\parskip}{0.55em}
        \setlength{\parindent}{0pt}
    },
    drop fuzzy shadow
}
\newtcolorbox{promptbox}[1][]{
  breakable,
  enhanced,
  colback=gray!8,
  colframe=gray!50,
  fonttitle=\bfseries\small,
  title={#1},
  left=6pt, right=6pt, top=4pt, bottom=4pt,
  arc=3pt,
}

\usepackage{inconsolata}
\usepackage{multirow}                 
\usepackage{multicol}                
\usepackage{multirow}               
\usepackage{booktabs}
\usepackage{enumitem}
\usepackage{graphicx}
\usepackage{xcolor}
\usepackage{colortbl}
\usepackage{subcaption}
\usepackage{pgfplots}
\pgfplotsset{compat=1.18}

\usepackage{tikz}
\usetikzlibrary{shapes.geometric}

\title{EntangleCodec: A Unified Discrete Audio Tokenizer via Semantic-Acoustic Entanglement}

\author{
\textbf{Hui Li}\thanks{~Equal contribution.}, \quad
\textbf{Yangfan Gao}\footnotemark[1], \quad
\textbf{Junlin Shang}, \quad
\textbf{Changhao Jiang}, \\
\textbf{Tao Gui}, \quad
\textbf{Qi Zhang}\thanks{~Corresponding authors.}, \quad
\textbf{Xuanjing Huang} \\
Fudan University \\
\texttt{hui\_li25@m.fudan.edu.cn, ~qzhang@fudan.edu.cn}
}
\usepackage{xcolor}
\usepackage{colortbl}
\usepackage{soul}

\definecolor{bestrow}{RGB}{248, 255, 235}
\definecolor{ours}{RGB}{230, 245, 255}
\definecolor{deltagreen}{RGB}{34, 139, 34}
\definecolor{deltared}{RGB}{200, 50, 50}

\begin{document}
\maketitle
\begin{abstract}
Audio tokenizers bridge continuous audio and Audio Language Models (ALMs), yet existing methods often trade off understanding and generation. Reconstruction-oriented codecs lack rich semantics, while semantic-aware tokenizers typically use separate semantic and acoustic streams, introducing redundancy or misalignment.

We propose \textbf{EntangleCodec}, a unified discrete audio tokenizer that learns caption-aligned semantic-acoustic representations before quantization. Rich audio captions enable a compact token stream to capture linguistic, speaker, prosodic, emotional, and acoustic information beyond ASR transcripts. A flow-matching diffusion decoder further supports high-quality reconstruction across speech, music, and general audio.

EntangleCodec achieves reconstruction quality competitive with specialized codecs, outperforms all codec-based baselines on audio understanding by up to \textbf{+7.4\%} on MMAR, and supports both TTS and TTA generation in a unified framework. Furthermore, EntangleCodec-based audio language models demonstrate strong scaling behavior: even at \textit{0.6B} parameters, the model surpasses specialized continuous-representation LLMs with over \textit{13B} parameters across three benchmarks using \textbf{22$\times$} fewer parameters; scaling to \textit{8B} further establishes new state-of-the-art results on MMAR, highlighting that representation quality is as critical as model scale in audio language modeling.
Code and model weights are available at 
\url{https://github.com/luckyerr/EntangleCodec}.
\end{abstract}

\section{Introduction}

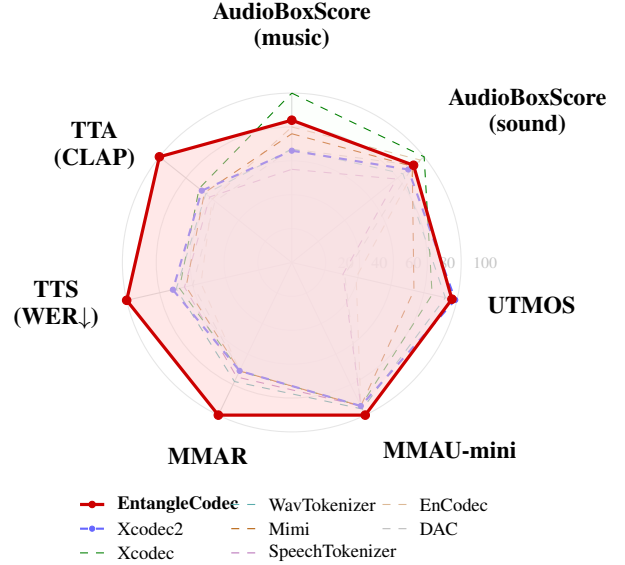
\begin{figure}[t]
\centering
\begin{tikzpicture}[scale=0.8]

\def\maxR{2.8cm}

\begin{scope}[shift={(0, 0)}]

\foreach \r/\lbl in {0.5/20, 1.0/40, 1.5/60, 2.0/80, 2.5/100}{
    \draw[gray!20, thin] (0,0) circle (\r*\maxR/2.5);
    \node[gray!45, font=\tiny, anchor=west] at (\r*\maxR/2.5+0.03cm, 0) {\lbl};
}

\foreach \angle in {90, 38.6, -12.9, -64.3, -115.7, -167.1, 141.4}{
    \draw[gray!35, thin] (0,0) -- (\angle:\maxR);
}

\draw[gray!55, thin, dashed, fill=gray!5, fill opacity=0.2]
    (90:{0.67*\maxR}) -- (38.6:{0.92*\maxR}) --
    (-12.9:{0.32*\maxR}) -- (-64.3:{0.94*\maxR}) --
    (-115.7:{0.71*\maxR}) -- (-167.1:{0.60*\maxR}) --
    (141.4:{0.58*\maxR}) -- cycle;

\draw[brown!55, thin, dashed, fill=brown!5, fill opacity=0.2]
    (90:{0.80*\maxR}) -- (38.6:{0.97*\maxR}) --
    (-12.9:{0.39*\maxR}) -- (-64.3:{0.94*\maxR}) --
    (-115.7:{0.71*\maxR}) -- (-167.1:{0.55*\maxR}) --
    (141.4:{0.60*\maxR}) -- cycle;

\draw[teal!65, thin, dashed, fill=teal!5, fill opacity=0.2]
    (90:{0.67*\maxR}) -- (38.6:{0.84*\maxR}) --
    (-12.9:{0.93*\maxR}) -- (-64.3:{0.96*\maxR}) --
    (-115.7:{0.78*\maxR}) -- (-167.1:{0.70*\maxR}) --
    (141.4:{0.65*\maxR}) -- cycle;

\draw[violet!55, thin, dashed, fill=violet!5, fill opacity=0.2]
    (90:{0.55*\maxR}) -- (38.6:{0.79*\maxR}) --
    (-12.9:{0.31*\maxR}) -- (-64.3:{0.94*\maxR}) --
    (-115.7:{0.75*\maxR}) -- (-167.1:{0.65*\maxR}) --
    (141.4:{0.62*\maxR}) -- cycle;

\draw[green!50!black, thin, dashed, fill=green!5, fill opacity=0.2]
    (90:{1.00*\maxR}) -- (38.6:{1.00*\maxR}) --
    (-12.9:{0.85*\maxR}) -- (-64.3:{0.94*\maxR}) --
    (-115.7:{0.71*\maxR}) -- (-167.1:{0.68*\maxR}) --
    (141.4:{0.70*\maxR}) -- cycle;

\draw[orange!70!black, thin, dashed, fill=orange!5, fill opacity=0.2]
    (90:{0.76*\maxR}) -- (38.6:{0.91*\maxR}) --
    (-12.9:{0.74*\maxR}) -- (-64.3:{0.94*\maxR}) --
    (-115.7:{0.71*\maxR}) -- (-167.1:{0.64*\maxR}) --
    (141.4:{0.66*\maxR}) -- cycle;

\draw[blue!60, thick, dashed, fill=blue!8, fill opacity=0.35]
    (90:{0.66*\maxR}) -- (38.6:{0.88*\maxR}) --
    (-12.9:{0.99*\maxR}) -- (-64.3:{0.94*\maxR}) --
    (-115.7:{0.71*\maxR}) -- (-167.1:{0.72*\maxR}) --
    (141.4:{0.68*\maxR}) -- cycle;
\foreach \angle/\val in {90/0.66, 38.6/0.88, -12.9/0.99, -64.3/0.94, -115.7/0.71, -167.1/0.72, 141.4/0.68}{
    \filldraw[blue!60] (\angle:{\val*\maxR}) circle (1.2pt);
}

\draw[red!80!black, very thick, fill=red!20, fill opacity=0.45]
    (90:{0.84*\maxR}) -- (38.6:{0.92*\maxR}) --
    (-12.9:{0.97*\maxR}) -- (-64.3:{1.00*\maxR}) --
    (-115.7:{1.00*\maxR}) -- (-167.1:{1.00*\maxR}) --
    (141.4:{1.00*\maxR}) -- cycle;
\foreach \angle/\val in {90/0.84, 38.6/0.92, -12.9/0.97, -64.3/1.00, -115.7/1.00, -167.1/1.00, 141.4/1.00}{
    \filldraw[red!80!black] (\angle:{\val*\maxR}) circle (2pt);
}

\node[font=\footnotesize\bfseries, anchor=south, align=center]
    at (90:{\maxR+0.55cm}) {AudioBoxScore\\(music)};
\node[font=\footnotesize\bfseries, anchor=south west, align=center]
    at (38.6:{\maxR+0.3cm}) {AudioBoxScore\\(sound)};
\node[font=\footnotesize\bfseries, anchor=west]
    at (-12.9:{\maxR+0.35cm}) {UTMOS};
\node[font=\footnotesize\bfseries, anchor=north west]
    at (-64.3:{\maxR+0.3cm}) {MMAU-mini};
\node[font=\footnotesize\bfseries, anchor=north, align=center]
    at (-115.7:{\maxR+0.4cm}) {MMAR};
\node[font=\footnotesize\bfseries, anchor=east, align=center]
    at (-167.1:{\maxR+0.3cm}) {TTS\\(WER$\downarrow$)};
\node[font=\footnotesize\bfseries, anchor=east, align=center]
    at (141.4:{\maxR+0.3cm}) {TTA\\(CLAP)};

\end{scope}

\begin{scope}[shift={(-3.5cm, -4.0cm)}]

    \draw[red!80!black, very thick] (0,0) -- (0.4cm,0);
    \filldraw[red!80!black] (0.2cm,0) circle (1.5pt);
    \node[anchor=west, font=\scriptsize\bfseries] at (0.45cm,0) {EntangleCodec};

    \draw[blue!60, thick, dashed] (0,-0.40cm) -- (0.4cm,-0.40cm);
    \filldraw[blue!60] (0.2cm,-0.40cm) circle (1.2pt);
    \node[anchor=west, font=\scriptsize] at (0.45cm,-0.40cm) {Xcodec2};

    \draw[green!50!black, thin, dashed] (0,-0.80cm) -- (0.4cm,-0.80cm);
    \node[anchor=west, font=\scriptsize] at (0.45cm,-0.80cm) {Xcodec};

    \draw[teal!65, thin, dashed] (2.5cm,0) -- (2.9cm,0);
    \node[anchor=west, font=\scriptsize] at (2.95cm,0) {WavTokenizer};

    \draw[orange!70!black, thin, dashed] (2.5cm,-0.40cm) -- (2.9cm,-0.40cm);
    \node[anchor=west, font=\scriptsize] at (2.95cm,-0.40cm) {Mimi};

    \draw[violet!55, thin, dashed] (2.5cm,-0.80cm) -- (2.9cm,-0.80cm);
    \node[anchor=west, font=\scriptsize] at (2.95cm,-0.80cm) {SpeechTokenizer};

    \draw[brown!55, thin, dashed] (5.0cm,0) -- (5.4cm,0);
    \node[anchor=west, font=\scriptsize] at (5.45cm,0) {EnCodec};

    \draw[gray!55, thin, dashed] (5.0cm,-0.40cm) -- (5.4cm,-0.40cm);
    \node[anchor=west, font=\scriptsize] at (5.45cm,-0.40cm) {DAC};

\end{scope}

\end{tikzpicture}
\caption{Comprehensive tokenizer comparison across seven dimensions (normalized 0--100; TTS WER inverted so higher = better).Reconstruction: AudioBoxScore (music/sound), UTMOS; Understanding: MMAU-mini, MMAR; Generation: TTS (WER$\downarrow$), TTA (CLAP).}
\vspace{-1em}
\label{fig:radar}
\end{figure}

The recent success of large language models has motivated growing interest in Audio Language Models (ALMs), spanning speech, music, and general audio understanding and generation. A central challenge is how to represent audio as discrete tokens that are both semantically informative and acoustically faithful. Existing audio tokenizers often favor one side of this trade-off: reconstruction-oriented codecs preserve fine acoustic detail but provide limited semantic information, while semantic-aware tokenizers introduce additional semantic streams that require fusion with acoustic representations. Moreover, semantic supervision is often limited to ASR transcripts, ignoring non-textual cues such as speaker identity, emotion, prosody, acoustic scenes, and musical structure.

We propose \textbf{EntangleCodec}, a unified discrete audio tokenizer that learns caption-aligned semantic-acoustic representations before quantization. Instead of separating semantic and acoustic information into independent streams, EntangleCodec integrates them within a shared pre-quantization representation, reducing the need for late fusion. To enrich semantic supervision, we align audio with rich captions rather than ASR transcripts through contrastive learning, enabling the resulting tokens to capture linguistic content as well as speaker, affective, prosodic, and environmental cues. A flow-matching diffusion decoder then reconstructs high-quality audio from the discrete tokens, allowing the same tokenizer to support TTS, TTA, and audio QA without task-specific architectural changes.

\paragraph{Key empirical findings.}
EntangleCodec achieves reconstruction quality comparable to specialized codecs despite its unified design (UTMOS: 3.96 vs.\ 4.02). For audio understanding, it outperforms all codec-based baselines by up to \textbf{+7.4\%} on MMAR. Furthermore, a \textit{0.6B} EntangleCodec-based language model surpasses specialized continuous-representation models with over \textit{13B} parameters across MMAR, MMAU-mini, and MMAU, using \textbf{22$\times$} fewer parameters; scaling to \textit{8B} further establishes new state-of-the-art results on MMAR (\textbf{42.6\%}). These results highlight that tokenizer representation quality is as critical as model scale in audio language modeling.

\noindent Our main contributions are:
\begin{itemize}[leftmargin=*, itemsep=2pt]
    \item \textbf{EntangleCodec}, a unified codec-based tokenizer that learns semantic-acoustic entangled representations before quantization, reducing the need for dual-stream separation and late fusion while maintaining competitive reconstruction quality.

    \item \textbf{Caption-aligned discrete representation learning}, which goes beyond ASR-only supervision by contrastively aligning audio representations with rich captions covering linguistic content, speaker identity, emotion, prosody, and acoustic scenes, leading to more expressive discrete audio tokens.

    \item \textbf{A unified audio language modeling framework} for TTS, TTA, and audio QA, requiring no task-specific tokenizer or architectural modification and demonstrating the broad applicability of EntangleCodec across both generation and understanding tasks.
\end{itemize}

\begin{figure*}[h]
    \centering
    \includegraphics[width=1.0\textwidth]{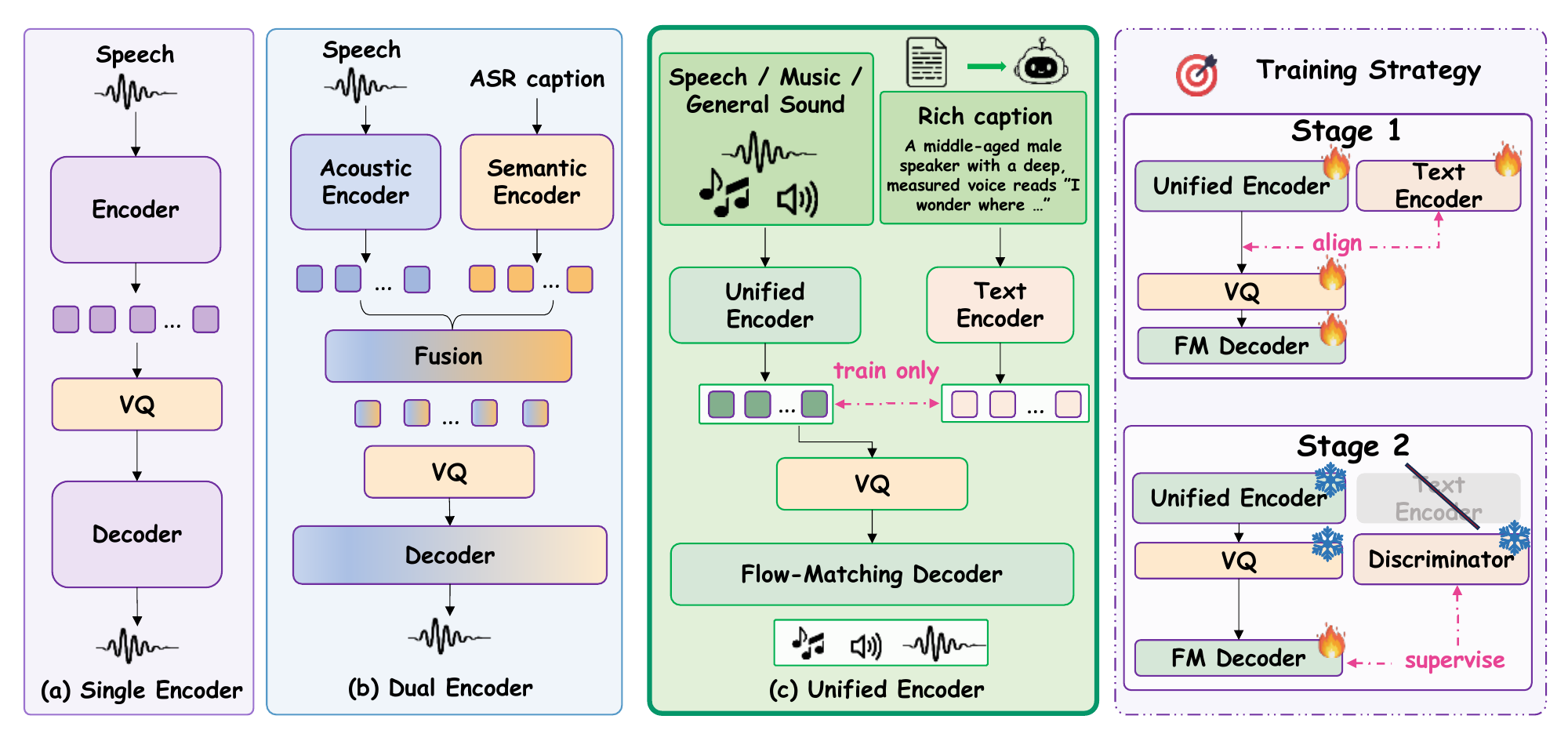}
\caption{
\textbf{Overview of EntangleCodec.}
Compared with single-encoder acoustic codecs and dual-encoder semantic-acoustic codecs, EntangleCodec learns caption-aligned unified tokens that support both reconstruction and downstream audio-language tasks.
The right panel shows the two-stage training strategy: joint codec learning with rich captions, followed by decoder refinement.
}
\label{fig:label}
\end{figure*}

\section{Related Work}
\subsection{Audio Codec and Tokenization}
Neural audio codecs provide the foundation for discrete audio language modeling. Early codecs such as SoundStream~\cite{soundstream} and EnCodec~\cite{encodec} utilized residual vector quantization (RVQ) for high-fidelity reconstruction at low bitrates, while DAC~\cite{dac} further improved codebook utilization and reconstruction quality. Recent tokenizers focus more explicitly on language-model compatibility: WavTokenizer~\cite{wavtokenizer} reduces token rates while preserving perceptual quality, Mimi~\cite{mimi} supports streaming-friendly tokenization, and SpeechTokenizer~\cite{speechtokenizer}, XCodec~\cite{xcodec}, and XCodec2~\cite{xcodec2} introduce semantic-aware designs for speech generation and audio modeling.

Most existing codecs follow one of two routes. Reconstruction-oriented codecs use a single acoustic encoder but provide limited semantic information for downstream reasoning. Semantic-aware codecs often introduce separate semantic and acoustic streams, which improves task relevance but requires additional fusion between independently learned representations. In contrast, EntangleCodec learns a caption-aligned semantic-acoustic representation before quantization, aiming to preserve both reconstruction fidelity and semantic richness within a unified discrete token stream.

\subsection{Audio-Language Models for Understanding}

Audio-language models extend LLMs to speech, music, and general audio understanding. LTU~\cite{ltu} and Audio Flamingo~\cite{audioflamingo} connect audio encoders with language models for audio question answering, while MU-LLaMA~\cite{mullama} and GAMA~\cite{gama} target music understanding and general audio reasoning. SALMONN~\cite{SALMONN} combines speech and audio encoders to handle diverse audio tasks, and Qwen2-Audio~\cite{qwen2audio} further improves performance through stronger audio encoders.

These systems typically rely on continuous audio representations produced by dedicated encoders. Such representations are effective for understanding, but they are not naturally suited to autoregressive audio generation. EntangleCodec instead provides discrete audio tokens that can be consumed by LLMs, enabling audio understanding and generation to share the same token interface.

\subsection{Text-Conditioned Audio Generation}

Text-conditioned audio generation includes text-to-speech (TTS) and text-to-audio (TTA). Autoregressive models such as AudioLM~\cite{audiolm} and VALL-E~\cite{valle} demonstrate the effectiveness of predicting discrete audio tokens for speech generation, while TTA models often rely on text-audio alignment signals such as CLAP~\cite{clap} to synthesize audio from natural language descriptions. Strong TTS systems such as NaturalSpeech3~\cite{naturalspeech3} and CosyVoice~\cite{cosyvoice} achieve high-quality speech synthesis, whereas general TTA systems focus on environmental sound, music, and acoustic scenes.

Despite this progress, TTS and TTA are usually treated with separate modeling pipelines, tokenizers, or conditioning mechanisms. EntangleCodec provides a unified discrete representation for both speech and general audio, allowing TTS and TTA to be formulated as text-conditioned token prediction tasks within the same audio language modeling framework.

\section{Methods}

\subsection{Overview}
EntangleCodec aims to build a unified discrete audio representation that simultaneously supports both understanding and generation tasks. As illustrated in Figure~\ref{fig:label}, the overall architecture consists of three core components: a unified encoder, a discrete quantizer, and a diffusion-based decoder. The fundamental distinction from existing audio tokenizers lies in two key design decisions: first, employing a single shared encoder rather than dual encoders to jointly model semantic and acoustic information in the same representation space; second, achieving semantic alignment through rich audio captions rather than ASR transcripts, encoding multimodal semantics beyond text.

\subsection{Unified Encoder}

\textbf{Unified Modeling.}
To obtain a discrete representation suitable for both audio understanding and generation, EntangleCodec integrates semantic and acoustic cues within the same pre-quantization representation. Unlike dual-encoder tokenizers that fuse separately extracted semantic and acoustic features, our encoder directly maps the input audio into a unified semantic-acoustic feature sequence, which is subsequently quantized into discrete tokens.

Given a Mel-spectrogram $\mathbf{M} \in \mathbb{R}^{128 \times T_m}$, we first project it into a $D_{enc}$-dimensional sequence:
\begin{equation}
    \mathbf{M}' = \mathrm{Linear}(\mathbf{M}^{\top}) \in \mathbb{R}^{T_m \times D_{enc}}.
\end{equation}
After prepending a learnable \texttt{[CLS]} token $\mathbf{c} \in \mathbb{R}^{D}$, the encoder input becomes
\begin{equation}
    \mathbf{E} = [\mathbf{c}; \mathbf{M}'] \in \mathbb{R}^{(T_m+1) \times D_{enc}}.
\end{equation}
This sequence is processed by the Transformer-based encoder:
\begin{equation}
    \mathbf{H}' = \mathrm{Encoder}(\mathbf{E}) \in \mathbb{R}^{(T_m+1) \times D_{enc}}.
\end{equation}
We discard the \texttt{[CLS]} token and use the remaining frame-level outputs
$\mathbf{H} \in \mathbb{R}^{T_m \times D_{enc}}$ as the unified semantic-acoustic representation for subsequent quantization. By performing semantic-acoustic integration before quantization, EntangleCodec avoids an explicit late-fusion module and provides a shared representation space for both reconstruction and downstream audio understanding.

\paragraph{Rich Semantic Alignment.}
To enrich the semantic content of the unified audio representation, EntangleCodec aligns audio with rich captions rather than ASR transcripts. Unlike transcripts that mainly preserve linguistic content, rich captions additionally describe speaker attributes, emotion, prosody, acoustic scenes, and sound events, providing a broader supervision signal for semantic-acoustic representation learning. Notably, these rich captions are automatically generated by a large language model (LLM), which comprehensively characterizes the multi-dimensional semantic attributes of audio in natural language, thereby providing the encoder with a substantially richer supervision signal than conventional ASR transcripts, details in Appendix~\ref{sec:Caption}.

We implement this alignment through audio-text contrastive learning. Given the encoder output $\mathbf{H}'$, we take the class-token representation $\mathbf{h}_{\mathrm{cls}} = \mathbf{H}'_0$ as the global audio embedding and project it into a $D_{align}$-dimensional space:
\begin{align}
    \mathbf{a} &= \mathrm{Proj}_{\mathrm{audio}}(\mathbf{h}_{\mathrm{cls}}) \in \mathbb{R}^{D_{align}}, \\
    \mathbf{t}_{\mathrm{emb}} &= \mathrm{TextEncoder}(\mathbf{t}) \in \mathbb{R}^{D_{align}},
\end{align}
where $\mathbf{t}$ denotes the rich caption generated by the LLM, and $\mathrm{TextEncoder}(\cdot)$ is a 12-layer Transformer text encoder. The audio and text embeddings are then aligned using a CLIP-style contrastive loss, encouraging the unified encoder to encode semantic cues beyond transcript-level content while retaining the acoustic information required for reconstruction.

\textbf{Pre-quantization Processing.}
The unified representation $\mathbf{H}$ is projected to $D_{quant}$ dimensions via a two-layer MLP and is L2-normalized to satisfy the constraints of single-codebook vector quantization:
\begin{equation}
    \mathbf{h} = \mathrm{MLP}_{\mathrm{VQ}}(\mathbf{H}) \in \mathbb{R}^{T_m \times D_{quant}},
    \quad
    \tilde{\mathbf{h}} = \frac{\mathbf{h}}{\|\mathbf{h}\|_2}.
\end{equation}
The normalized representation $\tilde{\mathbf{h}}$ is then fed into a single-codebook vector quantizer for discretization, producing the final token sequence.

\subsection{Diffusion-based Decoder}
The diffusion decoder reconstructs Mel-spectrograms from the quantized
representation $\mathbf{Z}_q$ using Flow Matching~\cite{flowmatching}.
The decoder is implemented as a Llama-style Transformer flow predictor,
taking the noisy Mel-spectrogram $\mathbf{x}_t$, a time-step embedding,
and $\mathbf{Z}_q$ as a conditioning prefix as input, and predicting
the velocity field:
\begin{equation}
    \mathbf{v}_\theta(\mathbf{x}_t, t, \mathbf{Z}_q) \in \mathbb{R}^{T_m \times 128}.
\end{equation}

\subsection{Training Objectives and Inference}

EntangleCodec adopts a two-stage training strategy. The first stage jointly
trains the audio encoder, text encoder, quantizer, and diffusion decoder,
enabling the encoder to capture rich multimodal semantics through contrastive
learning. The second stage freezes the encoder and quantizer, discards the
text encoder, and trains only the decoder to refine reconstruction quality.

\textbf{Stage 1: Semantic Learning and Joint Optimization.}
The total loss consists of three components:
\begin{equation}
    \mathcal{L}_{\text{stage1}} = \mathcal{L}_{\text{flow}} + \mathcal{L}_{\text{contrast}} + \mathcal{L}_{\text{vq}}
\end{equation}
\textbf{Flow Matching Loss} $\mathcal{L}_{\text{flow}}$ is based on the
Rectified Flow objective. Given an interpolated sample
$\mathbf{x}_t = (1-(1-\sigma_{\min})t)\mathbf{z} + t\mathbf{x}$,
the decoder is trained to predict the velocity field:
\begin{equation}
    \mathcal{L}_{\text{flow}} = \mathbb{E}_{t,\mathbf{x},\mathbf{z}}
    \left[\|\mathbf{v}_\theta(\mathbf{x}_t, t, \mathbf{Z}_q) -
    (\mathbf{x} - (1-\sigma_{\min})\mathbf{z})\|_1\right]
\end{equation}
\textbf{Contrastive Loss} $\mathcal{L}_{\text{contrast}}$ aligns audio with
LLM-generated captions via a CLIP-style bidirectional cross-entropy loss,
enabling the encoder to learn rich semantics including speaker characteristics,
emotion, and prosody:
\begin{equation}
    \mathcal{L}_{\text{contrast}} = \frac{1}{2}\left(\mathcal{L}_{\text{a2t}} + \mathcal{L}_{\text{t2a}}\right)
\end{equation}
\textbf{VQ Commitment Loss} $\mathcal{L}_{\text{vq}}$ stabilizes quantization
and prevents codebook collapse:
\begin{equation}
    \mathcal{L}_{\text{vq}} = \|\text{sg}(\tilde{\mathbf{h}}) - \mathbf{z}_q\|_2^2
    + \beta\|\tilde{\mathbf{h}} - \text{sg}(\mathbf{z}_q)\|_2^2
\end{equation}

\textbf{Stage 2: Refining Reconstruction Quality.}
The encoder and quantizer are frozen, and only the decoder is updated using the reconstruction-oriented flow and VQ objectives together with an adversarial loss (Appendix~\ref{app:gan}). This stage improves perceptual reconstruction quality while preserving the semantic-acoustic representation learned in Stage~1.

The two stages are trained for 500k and 200k steps respectively, with 10\%
conditional dropout applied throughout to support classifier-free guidance.

\textbf{Inference.}
For \textbf{audio reconstruction}, $\mathbf{Z}_q$ is
obtained by encoding and quantizing the input audio. For \textbf{conditional
generation} (e.g., TTS, TTA), an autoregressive LLM predicts token indices from
the text condition, and the corresponding codebook embeddings serve as $\mathbf{Z}_q$.
Both modes share the same decoder and differ only in the source of discrete tokens.
Sampling is performed with a single Euler step, with optional classifier-free guidance:
\begin{equation}
    \tilde{\mathbf{v}}_\theta
    = (1+\gamma)\,\mathbf{v}_\theta(\mathbf{x}_t, t, \mathbf{Z}_q)
    - \gamma\,\mathbf{v}_\theta(\mathbf{x}_t, t, \varnothing),
\end{equation}
where $\varnothing$ denotes the null condition and $\gamma=1.0$ by default.
The resulting Mel-spectrogram is converted to a waveform using the Vocos vocoder.

\section{Experiments}
\subsection{Experimental Setup}

\begin{table*}[ht]
  \centering
  \setlength{\tabcolsep}{4pt}
  \begin{tabular}{lc|cccc|cc}
    \toprule
    \multirow{2}{*}{\textbf{Model}} & 
    \multirow{2}{*}{\textbf{T./L.}} & 
    \multicolumn{4}{c|}{\textbf{Speech}} & 
    \textbf{Sound} & \textbf{Music} \\
    \cmidrule(lr){3-6}\cmidrule(lr){7-7}\cmidrule(lr){8-8}
    & & \textbf{UTMOS}$\uparrow$ & \textbf{F1}$\uparrow$ & 
      \textbf{STOI}$\uparrow$ & \textbf{SIM}$\uparrow$ & 
      \textbf{AudioBoxScore} & \textbf{AudioBoxScore} \\
    \midrule
    GT              & /    & 4.08 & 0.98 & 1.00 & 1.00 & 3.64/5.16/4.30/5.84 & 5.93/6.43/5.24/6.59 \\
    \midrule
    DAC-Codec       & 50/1 & 1.30 & 0.97 & 0.62 & 0.25 & 3.16/3.99/4.25/5.07 & 3.58/3.97/5.18/4.61 \\
    EnCodec         & 75/1 & 1.57 & 0.92 & 0.77 & 0.25 & 3.39/4.86/3.46/5.63 & 4.92/5.32/4.74/5.52 \\
    WavTokenizer    & 75/1 & 3.79 & \underline{0.98} & \textbf{0.90} & 0.65 & 2.84/3.99/3.04/5.07 & 3.59/4.39/4.18/5.16 \\
    SpeechTokenizer & 50/1 & 1.27 & 0.97 & 0.64 & 0.17 & 2.48/4.10/2.43/5.20 & 2.34/3.85/2.88/5.00 \\
    Xcodec          & 50/1 & 3.42 & 0.97 & 0.85 & 0.48 & \textbf{3.48/4.66/4.11/5.62} & \textbf{6.25/6.75/5.82/6.92} \\
    Mimi            & 50/4 & 3.03 & 0.97 & 0.85 & 0.50 & 3.14/4.25/3.67/5.24 & 4.12/4.81/5.26/5.37 \\
    Xcodec2         & 50/1 & \textbf{4.02} & \underline{0.98} & \underline{0.88} & \textbf{0.76} & 3.13/4.14/3.08/5.27 & 3.26/4.26/4.08/5.45 \\
    WavTokenizer    & 40/1 & 3.58 & 0.97 & 0.85 & 0.48 & 3.11/4.30/3.58/5.19 & 4.78/5.34/5.29/5.61 \\
    \midrule
    \rowcolor{ours}
    \textbf{EntangleCodec} & \textbf{50/1} 
      & \underline{3.96} 
      & \textbf{0.98} 
      & 0.87 
      & \underline{0.66} 
      & \underline{3.12/4.50/3.36/5.45}
      & \underline{4.98/5.70/5.10/5.88} \\
    \bottomrule
  \end{tabular}
  \caption{Tokenizer reconstruction quality comparison. \textbf{Bold} = best result; 
           \underline{underline} = second best (GT excluded). 
           $\Delta$ denotes improvement of EntangleCodec over the second-best baseline.}
    \label{tab:reconstruction}
\end{table*}
\begin{table*}[ht]
  \centering
  \begin{tabular}{lc|cccc}
    \toprule
    \textbf{Model} & \textbf{TPS/VQ Layers} & 
    \textbf{MMAU-mini}$\uparrow$ & \textbf{MMAU}$\uparrow$ & 
    \textbf{MMAR}$\uparrow$ & \textbf{Average}$\uparrow$ \\
    \midrule
    WavTokenizer    & 75/1  & \underline{32.7} & 32.3 & \underline{26.9} & \underline{30.6} \\
    SpeechTokenizer & 400/8 & 31.9 & 31.6  & 25.9 & 29.8 \\
    Xcodec          & 50/1  & 32.0 & \underline{32.6}  & 24.2 & 29.6 \\
    Xcodec2         & 50/1  & 26.8  &25.7  & 24.6 & 25.7 \\
    XY-Tokenizer    & 100/8 & 31.6 & 30.1 & 24.9 & 28.9 \\
    \midrule
    \rowcolor{ours}
    \textbf{EntangleCodec (Ours)} & \textbf{50/1} 
      & \textbf{34.2} & \textbf{35.1} & \textbf{34.3} & \textbf{34.5} \\
    \midrule
    \rowcolor{bestrow}
    $\Delta$ vs.\ 2nd best & & 
      \textcolor{deltagreen}{$+$1.5} & 
      \textcolor{deltagreen}{$+$2.5} & 
      \textcolor{deltagreen}{$+$7.4} & 
      \textcolor{deltagreen}{$+$3.9} \\
    \bottomrule
  \end{tabular}
   \caption{Audio understanding performance. All discrete-tokenizer models share the same 
           Qwen3-0.6B backbone. \textbf{Bold} = best; \underline{underline} = second best.
           $\Delta$ denotes improvement over the second-best baseline.}
    \label{tab:understanding}
\end{table*}

\subsubsection{Datasets}

\paragraph{Training Data.}
We train EntangleCodec on a diverse corpus covering speech, music, and general audio, including LibriSpeech~\cite{librispeech}, MusicBench~\cite{musicbench}, AudioSet~\cite{audioset}, AudioCaps~\cite{audiocaps}, and WavCaps~\cite{wavcaps}. Downstream audio language models are trained on task-specific instruction and generation data for audio understanding, TTS, and TTA. Detailed dataset statistics are provided in Appendix~\ref{sec:data}.

\paragraph{Evaluation Data.}
We evaluate reconstruction on held-out speech, music, and general-audio test sets, audio understanding on MMAR~\cite{mmar}, MMAU-mini~\cite{mmau}, and MMAU~\cite{mmau}, and audio generation on standard TTS and TTA benchmarks. Detailed evaluation sets are listed in Appendix~\ref{sec:data}.

\begin{table*}[ht]
  \centering
  \begin{tabular}{lc|cc|cc}
    \toprule
    \multirow{2}{*}{\textbf{Model}} & \multirow{2}{*}{\textbf{TPS/VQ Layers}} &
    \multicolumn{2}{c|}{\textbf{TTS (Speech)}} &
    \multicolumn{2}{c}{\textbf{TTA (Sound)}} \\
    \cmidrule(lr){3-4}\cmidrule(lr){5-6}
    & & \textbf{WER}$\downarrow$ & \textbf{UTMOS}$\uparrow$
      & \textbf{AudioBox Score}$\uparrow$ & \textbf{CLAP}$\uparrow$ \\
    \midrule
    WavTokenizer & 75/1  & 17.2 & 1.29 & 2.85/3.39/2.48/5.20 & 0.01 \\
    Xcodec       & 50/1  & 25.6 & 2.95 & \textbf{3.58/5.06/4.72/6.01} & 0.03 \\
    Xcodec2      & 50/1  & 23.2 & 1.53 & 3.07/5.11/2.51/4.37 & 0.02 \\
    XY-Tokenizer & 100/8 & \underline{12.1} & \underline{2.95} & \underline{3.17/4.81/3.02/6.01} & \underline{0.04} \\
    \midrule
    \rowcolor{ours}
    \textbf{EntangleCodec} & 50/1 & \textbf{9.8} & \textbf{3.89} & 3.29/4.85/3.19/5.75 & \textbf{0.17} \\
    \midrule
    \rowcolor{bestrow}
    $\Delta$ vs.\ 2nd best & &
      \textcolor{deltagreen}{$-2.3$} &
      \textcolor{deltagreen}{$+0.94$} &
      \textcolor{deltared}{$-0.57$} &
      \textcolor{deltagreen}{$+0.13$} \\
    \bottomrule
  \end{tabular}
  \caption{Audio generation performance. All discrete-tokenizer models share the same
           Qwen3-0.6B backbone. \textbf{Bold} = best; \underline{underline} = second best.
           $\Delta$ denotes improvement over the second-best baseline.
           TTS WER: lower is better (shown inverted for $\Delta$).}
    \label{tab:generation}
\end{table*}

\subsubsection{Implementation Details}

EntangleCodec is trained in two stages: Stage~1 jointly learns the tokenizer and decoder with reconstruction, contrastive, and VQ losses; Stage~2 freezes the tokenizer and refines the decoder for reconstruction. After training, EntangleCodec is frozen and used to train Qwen3-series audio language models~\cite{qwen3} for understanding and generation. Detailed optimization settings and hyperparameters are provided in Appendix~\ref{app:hp}.
\subsubsection{Baselines}

We compare EntangleCodec along three dimensions. For reconstruction, we evaluate against representative neural audio codecs under matched compression settings using official pretrained checkpoints. For audio understanding, we use a codec-controlled setting where identical Qwen3-0.6B~\cite{qwen3} language models are trained on tokens from different codecs, isolating the effect of tokenization. We additionally compare with specialized continuous-representation audio LLMs using official checkpoints or reported results when available. For generation, we follow the same codec-controlled protocol for TTS and TTA. The full list of baselines is provided in Appendix~\ref{app:baseline}.

\begin{figure*}[t]
\centering

\pgfplotsset{
    scaling plot/.style={
        width=0.33\textwidth,
        height=5.5cm,
        xmin=0, xmax=15,
        ymin=10, ymax=60,
        xtick={0,2,4,6,8,10,12,14},
        ytick={10,15,20,25,30,35,40,45,50,55,60},
        grid=both,
        grid style={line width=0.3pt, draw=gray!20},
        major grid style={line width=0.4pt, draw=gray!35},
        tick label style={font=\footnotesize},
        label style={font=\footnotesize\bfseries},
        xlabel={\textbf{Parameters (B)}},
        clip=false,
    },
    scaling legend/.style={
        legend columns=-1,
        legend style={
            font=\tiny,
            draw=gray!40,
            fill=white,
            rounded corners=1pt,
            column sep=4pt,
            row sep=1pt,
            /tikz/every even column/.append style={column sep=6pt},
        }
    }
}

\begin{tikzpicture}
\begin{axis}[
    scaling plot,
    scaling legend,
    ylabel={\textbf{MMAU-mini (\%)}},
    title={\textbf{(a) MMAU-mini}},
    legend to name=scalinglegend,
]
\addplot[only marks, mark=*, mark size=3pt, color=gray!60, fill=gray!30]
    coordinates {(7,16.89)(7,17.68)(2.2,16.69)(7,31.90)(7,30.90)};
\addlegendentry{Other Baselines}

\addplot[only marks, mark=square*, mark size=3.5pt, color=orange!80!black, fill=orange!50]
    coordinates {(13,33.70)};
\addlegendentry{SALMONN (13B)}

\addplot[only marks, mark=triangle*, mark size=4.5pt, color=purple!70, fill=purple!40]
    coordinates {(8.4,49.20)};
\addlegendentry{Qwen2-Audio (8.4B)}

\addplot[only marks, mark=diamond*, mark size=4pt, color=blue!70, fill=blue!40]
    coordinates {(4,52.3)};
\addlegendentry{\textbf{EntangleCodec (4B)}}

\addplot[only marks, mark=pentagon*, mark size=4pt, color=teal!80!black, fill=teal!40]
    coordinates {(8,56.2)};
\addlegendentry{\textbf{EntangleCodec (8B)}}

\addplot[only marks, mark=star, mark size=7pt, color=red!80!black, fill=red!60, line width=1pt]
    coordinates {(0.6,34.2)};
\addlegendentry{\textbf{EntangleCodec (0.6B)}}

\node[font=\tiny, color=gray!70, anchor=north, yshift=-2pt] at (axis cs:7,16.89) {LTU};
\node[font=\tiny, color=gray!70, anchor=south, yshift=2pt] at (axis cs:7,17.68) {LTU AS};
\node[font=\tiny, color=gray!70, anchor=east, xshift=-3pt] at (axis cs:2.2,16.69) {Flamingo};
\node[font=\tiny, color=gray!70, anchor=north, yshift=-2pt] at (axis cs:7,31.90) {MU-LLaMA};
\node[font=\tiny, color=gray!70, anchor=south, yshift=2pt] at (axis cs:7,30.90) {GAMA};
\node[font=\tiny, color=orange!80!black, anchor=south, yshift=3pt] at (axis cs:13,33.70) {SALMONN};
\node[font=\tiny, color=purple!80, anchor=south, yshift=3pt] at (axis cs:8.4,49.20) {Qwen2-Audio};
\node[font=\tiny\bfseries, color=blue!70, anchor=west, xshift=4pt] at (axis cs:4,52.3) {\textbf{52.3}};
\node[font=\tiny\bfseries, color=teal!80!black, anchor=west, xshift=4pt] at (axis cs:8,56.2) {\textbf{56.2}};
\node[font=\tiny\bfseries, color=red!80!black, anchor=west, xshift=4pt] at (axis cs:0.6,34.2) {\textbf{34.2}};
\end{axis}
\end{tikzpicture}
\hfill
\begin{tikzpicture}
\begin{axis}[
    scaling plot,
    ylabel={\textbf{MMAU (\%)}},
    title={\textbf{(b) MMAU}},
]
\addplot[only marks, mark=*, mark size=3pt, color=gray!60, fill=gray!30]
    coordinates {(7,18.51)(7,18.90)(2.2,18.87)(7,30.66)(7,31.81)};
\addplot[only marks, mark=square*, mark size=3.5pt, color=orange!80!black, fill=orange!50]
    coordinates {(13,32.77)};
\addplot[only marks, mark=triangle*, mark size=4.5pt, color=purple!70, fill=purple!40]
    coordinates {(8.4,52.50)};
\addplot[only marks, mark=diamond*, mark size=4pt, color=blue!70, fill=blue!40]
    coordinates {(4,50.6)};
\addplot[only marks, mark=pentagon*, mark size=4pt, color=teal!80!black, fill=teal!40]
    coordinates {(8,52.6)};
\addplot[only marks, mark=star, mark size=7pt, color=red!80!black, fill=red!60, line width=1pt]
    coordinates {(0.6,35.1)};

\node[font=\tiny, color=gray!70, anchor=north, yshift=-2pt] at (axis cs:7,18.51) {LTU};
\node[font=\tiny, color=gray!70, anchor=south, yshift=2pt] at (axis cs:7,18.90) {LTU AS};
\node[font=\tiny, color=gray!70, anchor=east, xshift=-3pt] at (axis cs:2.2,18.87) {Flamingo};
\node[font=\tiny, color=gray!70, anchor=north, yshift=-2pt] at (axis cs:7,30.66) {MU-LLaMA};
\node[font=\tiny, color=gray!70, anchor=south, yshift=2pt] at (axis cs:7,31.81) {GAMA};
\node[font=\tiny, color=orange!80!black, anchor=south, yshift=3pt] at (axis cs:13,32.77) {SALMONN};
\node[font=\tiny, color=purple!80, anchor=south, yshift=3pt] at (axis cs:8.4,52.50) {Qwen2-Audio};
\node[font=\tiny\bfseries, color=blue!70, anchor=west, xshift=4pt] at (axis cs:4,50.6) {\textbf{50.6}};
\node[font=\tiny\bfseries, color=teal!80!black, anchor=south, yshift=3pt] at (axis cs:8,52.6) {\textbf{52.6 (\#1)}};
\node[font=\tiny\bfseries, color=red!80!black, anchor=west, xshift=4pt] at (axis cs:0.6,35.1) {\textbf{35.1}};
\end{axis}
\end{tikzpicture}
\hfill
\begin{tikzpicture}
\begin{axis}[
    scaling plot,
    ylabel={\textbf{MMAR (\%)}},
    title={\textbf{(c) MMAR}},
]
\addplot[only marks, mark=*, mark size=3pt, color=gray!60, fill=gray!30]
    coordinates {(7,19.20)(7,19.00)(2.2,26.60)(7,13.90)(7,26.40)};
\addplot[only marks, mark=square*, mark size=3.5pt, color=orange!80!black, fill=orange!50]
    coordinates {(13,33.20)};
\addplot[only marks, mark=triangle*, mark size=4.5pt, color=purple!70, fill=purple!40]
    coordinates {(8.4,30.00)};
\addplot[only marks, mark=diamond*, mark size=4pt, color=blue!70, fill=blue!40]
    coordinates {(4,41.8)};
\addplot[only marks, mark=pentagon*, mark size=4pt, color=teal!80!black, fill=teal!40]
    coordinates {(8,42.6)};
\addplot[only marks, mark=star, mark size=7pt, color=red!80!black, fill=red!60, line width=1pt]
    coordinates {(0.6,34.3)};

\node[font=\tiny, color=gray!70, anchor=north, yshift=-2pt] at (axis cs:7,19.20) {LTU};
\node[font=\tiny, color=gray!70, anchor=south, yshift=2pt] at (axis cs:7,19.00) {LTU AS};
\node[font=\tiny, color=gray!70, anchor=east, xshift=-3pt] at (axis cs:2.2,26.60) {Flamingo};
\node[font=\tiny, color=gray!70, anchor=north, yshift=-2pt] at (axis cs:7,13.90) {MU-LLaMA};
\node[font=\tiny, color=gray!70, anchor=south, yshift=2pt] at (axis cs:7,26.40) {GAMA};
\node[font=\tiny, color=orange!80!black, anchor=south, yshift=3pt] at (axis cs:13,33.20) {SALMONN};
\node[font=\tiny, color=purple!80, anchor=east, xshift=-3pt] at (axis cs:8.4,30.00) {Qwen2-Audio};
\node[font=\tiny\bfseries, color=blue!70, anchor=west, xshift=4pt] at (axis cs:4,41.8) {\textbf{41.8}};
\node[font=\tiny\bfseries, color=teal!80!black, anchor=west, xshift=4pt] at (axis cs:8,42.6) {\textbf{42.6 (\#1)}};
\node[font=\tiny\bfseries, color=red!80!black, anchor=west, xshift=4pt] at (axis cs:0.6,34.3) {\textbf{34.3}};
\draw[->, dashed, thick, color=teal!60]
    (axis cs:9,42.6) -- (axis cs:12,33.20)
    node[midway, above, font=\tiny, color=teal!70, sloped] {fewer params};
\end{axis}
\end{tikzpicture}

\vspace{2pt}

\ref{scalinglegend}

\caption{
Parameter efficiency across audio understanding benchmarks.
EntangleCodec scales consistently from 0.6B to 8B, achieving the best results on MMAU at 8B and on MMAR at both 4B and 8B, while outperforming larger specialized audio LLMs with fewer parameters.
}
\label{fig:scaling}
\end{figure*}
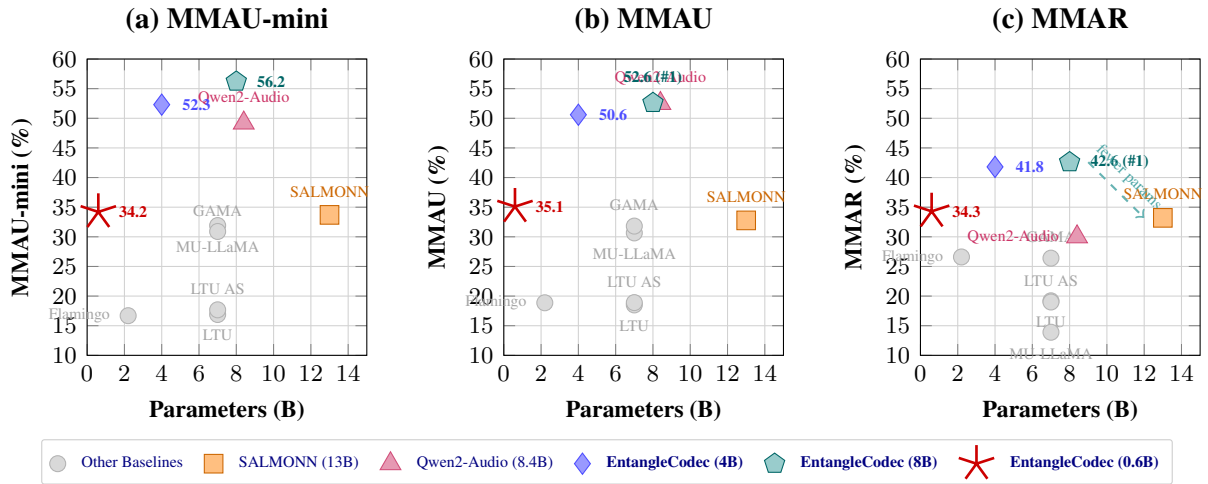

\begin{table*}[ht]
  \centering
  \setlength{\tabcolsep}{2pt}
  \renewcommand{\arraystretch}{1.2}
  \begin{tabular}{l c | c c c c c c}
    \toprule
    \multirow{2}{*}{\textbf{Configuration}} & 
    \multirow{2}{*}{\textbf{T./L.}} & 
    \multicolumn{4}{c}{\textbf{Speech}} & 
    \multicolumn{2}{c}{\textbf{AudioBox Score}} \\
    \cmidrule(lr){3-6}\cmidrule(lr){7-8}
    & & \textbf{UTMOS}$\uparrow$ & \textbf{F1}$\uparrow$ & 
      \textbf{STOI}$\uparrow$ & \textbf{SIM}$\uparrow$ & 
      \textbf{Sound} & \textbf{Music} \\
    \midrule
    \rowcolor{ours}
    \textbf{EntangleCodec} & \textbf{50/1} 
      & \textbf{3.96} & \textbf{0.98} & \textbf{0.87} & \textbf{0.66} 
      & \textbf{3.12/4.50/3.36/5.45} & \textbf{4.98/5.70/5.10/5.88} \\
    w/o Contrastive Loss & 50/1 
      & 3.04$^{\textcolor{deltared}{\downarrow\text{0.92}}}$
      & 0.97$^{\textcolor{deltared}{\downarrow\text{0.01}}}$
      & 0.82$^{\textcolor{deltared}{\downarrow\text{0.05}}}$
      & 0.50$^{\textcolor{deltared}{\downarrow\text{0.16}}}$
      & 2.89/4.21/2.77/5.34 & 4.10/5.29/4.31/5.63 \\
    w/o Rich Caption & 50/1 
      & 3.24$^{\textcolor{deltared}{\downarrow\text{0.72}}}$
      & 0.97$^{\textcolor{deltared}{\downarrow\text{0.01}}}$
      & 0.83$^{\textcolor{deltared}{\downarrow\text{0.04}}}$
      & 0.51$^{\textcolor{deltared}{\downarrow\text{0.15}}}$
      & 2.64/3.92/2.69/5.13 & 3.65/4.71/3.45/5.34 \\
    w/o Stage 2 & 50/1 
      & 3.40$^{\textcolor{deltared}{\downarrow\text{0.56}}}$
      & 0.98$^{\textcolor{deltagreen}{=}}$
      & 0.85$^{\textcolor{deltared}{\downarrow\text{0.02}}}$
      & 0.54$^{\textcolor{deltared}{\downarrow\text{0.12}}}$
      & 2.84/4.22/2.76/5.34 & 4.35/5.53/4.35/5.79 \\
    \bottomrule
  \end{tabular}
  \caption{Ablation study on key components of EntangleCodec. 
           \textbf{Bold} = best (full model); 
           superscript \textcolor{deltared}{$\downarrow$} indicates degradation from full model.}
  \label{tab:ablation}
\end{table*}


\subsubsection{Evaluation Metrics}

We evaluate reconstruction with UTMOS~\cite{utmos}, PESQ~\cite{pesq}, voiced/unvoiced F1, and the 4 metrics in AudioBox Aesthetics~\cite{audiobox} (AudioBoxScore); audio understanding with accuracy; TTS with WER and UTMOS; and TTA with CLAP Score~\cite{clap}. Lower WER is better, while higher values are better for all other metrics. Detailed metric definitions are provided in Appendix~\ref{app:metrics}.
\subsection{Results and Analysis}





\subsubsection{Finding 1: EntangleCodec Maintains Strong Reconstruction Quality}

\noindent\textbf{$\triangleright$} \textit{EntangleCodec achieves reconstruction quality comparable to specialized neural codecs while providing a unified semantic-acoustic representation.}

\paragraph{Speech Reconstruction.}
Our model achieves \textbf{3.96} UTMOS and \textbf{0.98} F1, approaching the best baselines of 4.02 and 0.98, respectively., and remains competitive on STOI and SIM. Learning a semantic-aware representation does not come at the cost of speech reconstruction quality.

\paragraph{Sound and Music Reconstruction.}
Our model ranks second on both sound (\textbf{4.11}) and music (\textbf{5.42}) AudioBoxScore, trailing only XCodec which lacks semantic alignment. EntangleCodec is the only tokenizer that consistently ranks among the top performers across all three audio domains, confirming that semantic-acoustic entanglement and reconstruction fidelity are complementary rather than conflicting objectives.

\subsubsection{Finding 2: Representation Quality Improves Audio Understanding}

\noindent\textbf{$\triangleright$} \textit{Under controlled comparisons, EntangleCodec produces more effective discrete tokens for audio understanding; the resulting small audio LLM also remains competitive with much larger specialized models.}

\paragraph{Codec-controlled comparison.}
Table~\ref{tab:understanding} compares different tokenizers under the same backbone and training setup, isolating the effect of the codec itself. EntangleCodec achieves \textbf{34.2\%} on MMAU-mini, \textbf{35.11\%} on MMAU, and \textbf{34.3\%} on MMAR, outperforming the best discrete codec baseline by $+$\textbf{1.5\%}, $+$\textbf{2.51\%}, and $+$\textbf{7.4\%}, respectively, establishing a new state-of-the-art among discrete tokenizer-based audio LLMs. The consistent improvements across all three benchmarks indicate that EntangleCodec tokens provide stronger semantic cues for downstream reasoning than reconstruction-oriented codec tokens.

\paragraph{Comparison with continuous-representation audio LLMs.}
Figure~\ref{fig:scaling} shows EntangleCodec-LLM against continuous-representation audio LLMs. At only \textbf{0.6B} parameters, EntangleCodec-LLM already surpasses all discrete-tokenizer baselines and outperforms several much larger continuous-representation models, including SALMONN-13B by \textbf{+0.3\%}, \textbf{+0.6\%}, and \textbf{+1.1\%} on MMAU-mini, MMAU, and MMAR, and GAMA-7B by \textbf{+3.1\%}, \textbf{+1.6\%}, and \textbf{+7.9\%}, despite using \textbf{22$\times$} fewer parameters than SALMONN. These results demonstrate that a well-designed discrete codec is not inherently a bottleneck for audio understanding --- with sufficiently rich token representations, discrete-tokenizer LLMs can match or exceed continuous-representation systems at a fraction of the model size.

\paragraph{Effect of scaling.}
As shown in Figure~\ref{fig:scaling}, EntangleCodec-based LLMs scale effectively from \textbf{0.6B} to \textbf{4B} and \textbf{8B} parameters, with consistent and substantial gains across all benchmarks. At 4B, the model achieves \textbf{52.3\%}, \textbf{50.6\%}, and \textbf{41.8\%} on MMAU-mini, MMAU, and MMAR, surpassing Qwen2-Audio-8.4B on MMAU-mini by \textbf{+3.1\%} despite fewer parameters. Scaling to 8B further improves performance to \textbf{56.2\%}, \textbf{52.6\%}, and \textbf{42.6\%}, achieving state-of-the-art results on MMAU-mini and matching Qwen2-Audio on MMAU while surpassing all baselines on MMAR. The strong scaling behavior indicates that EntangleCodec provides a solid representational foundation that continues to benefit from increased language model capacity, suggesting that the performance gap with continuous-representation systems will narrow further as model scale grows.

\subsubsection{Finding 3: Unified Tokens Support Audio Generation}
\noindent\textbf{$\triangleright$} \textit{The same EntangleCodec tokens can be used for TTS and TTA generation without introducing task-specific tokenizers or separate acoustic representations.}
\paragraph{Text-to-speech.}
Table~\ref{tab:generation} shows that EntangleCodec achieves a WER of \textbf{9.8\%} and a UTMOS of \textbf{3.89} on TTS, ranking \textbf{first} among all codec-controlled baselines on both metrics. Compared to the second-best baseline, WER is reduced by 2.3 points and UTMOS improves by $+0.94$, demonstrating that the proposed tokenizer preserves sufficient acoustic and prosodic information for intelligible and natural speech synthesis.
\paragraph{Text-to-audio.}
For TTA, EntangleCodec achieves a CLAP Score of \textbf{0.17}, outperforming the second-best baseline by $+0.13$---more than $4\times$ the score of XY-Tokenizer (0.04). While AudioBoxScore is slightly below Xcodec ($-0.57$ on average), Xcodec exhibits substantially degraded TTS performance (WER 25.6\%), suggesting that its acoustic fidelity comes at the cost of semantic alignment. The strong CLAP improvement is consistent with the understanding results: caption-aligned semantic-acoustic tokens provide richer conditioning signals for generating audio that matches textual descriptions.

\subsection{Ablation Studies}
\label{sec:ablation}

We isolate the contribution of three core design choices in EntangleCodec: the audio-text contrastive learning objective, the richness of semantic supervision, and the two-stage training strategy. All ablated variants use the same architecture (Table~\ref{tab:ablation}).

\paragraph{Effect of Contrastive Learning.}
Removing the contrastive loss (\textit{w/o contrast}) leads to the most pronounced degradation among the ablated variants: UTMOS drops by \textbf{0.92} (3.96 $\to$ 3.04) and SIM drops by \textbf{0.16} (0.66 $\to$ 0.50), with consistent declines in AudioBoxScore for both sound and music. These results show that contrastive alignment is not merely an auxiliary semantic objective; it substantially shapes the encoder representation. By aligning audio features with text-level semantics, the contrastive objective encourages more structured representations that remain compatible with high-quality reconstruction.

\paragraph{Effect of Rich Captions vs.\ ASR-Only Supervision.}
Replacing rich audio captions with ASR transcripts (\textit{w/o rich semantic}) yields a smaller but consistent degradation across all metrics. UTMOS drops by \textbf{0.72} and SIM by \textbf{0.15}, while AudioBoxScore for both sound and music also declines. The gap between \textit{w/o contrast} and \textit{w/o rich semantic} suggests that ASR-only contrastive alignment still provides useful semantic supervision, but rich captions offer additional information beyond transcript content, including speaker attributes, emotion, prosody, and acoustic scenes. This supports our hypothesis that ASR-only supervision provides an incomplete semantic signal for unified audio tokenization.


\section{Conclusion}
We presented \textbf{EntangleCodec}, a unified discrete audio tokenizer that learns caption-aligned semantic-acoustic representations for both audio understanding and generation. By combining rich caption alignment with a flow-matching decoder, EntangleCodec supports reconstruction, TTS, TTA, and audio QA without task-specific tokenizers or architectural changes. Experiments show  EntangleCodec achieves competitive reconstruction quality, improves codec-based audio understanding by up to \textbf{+7.4\%} on MMAR, and enables a \textit{0.6B} audio LLM to surpass specialized models exceeding \textit{13B} parameters. These results highlight the importance of representation quality in audio language modeling.

\section*{Limitations}
While EntangleCodec demonstrates strong performance across multiple tasks, several limitations warrant further exploration and improvement.

\textbf{Fine-grained Semantic Modeling.} Although EntangleCodec uses rich audio captions to encode multimodal semantics, the granularity of semantics remains limited by annotation quality. For subtle semantic features---such as nuanced emotional shifts, complex musical harmonies, or specific acoustic scene details---current captions may not fully capture them. Finer-grained semantic modeling may require more detailed annotations or self-supervised learning methods.

\textbf{Limited Exploration of Larger Scales.} Due to computational resource and data constraints, our scaling experiments are limited to 8B parameters and a moderate amount of training data. While the consistent gains from 0.6B to 8B suggest that EntangleCodec would continue to benefit from further scaling, experiments at larger scales (e.g., 30B or 70B) remain unexplored. Furthermore, as parameters scale up, reconstruction and understanding tasks may have different demands on model capacity, and the optimal parameter allocation and architectural balance between the two objectives may shift. We leave a more comprehensive scaling analysis, along with strategies for balancing reconstruction and understanding under larger-scale training, to future work as resources permit.

\section*{Acknowledgements}
The authors wish to thank the anonymous reviewers for their helpful comments. This work was partially funded by National Key R\&D Program of China No.2025ZD0215702, National Natural Science Foundation of China (No. 62576106, 62476061, 62376061).


\bibliography{custom}

\clearpage
\appendix
\section{Claim of the Usage of AI Assistants}
In preparing this manuscript, AI assistants were used only to improve the clarity, style, and readability of selected passages. They did not contribute to the study design, methodological development or implementation, data collection or analysis, or the generation of the manuscript’s primary scientific contributions. All substantive research decisions, analyses, interpretations, and conclusions remain the sole responsibility of the authors.
\section{Implementation Details}
\label{sec:impl}

\subsection{Audio Preprocessing}

All audio is resampled to 24\,kHz. Log-Mel spectrograms are extracted with 128 Mel bins, a Hann window of 1024 samples (42.7\,ms), a hop size of 480 samples (20\,ms), and an FFT size of 1024, yielding a frame rate of 50\,Hz. The log-Mel values are computed from power spectrograms and normalized to zero mean and unit variance per utterance using statistics estimated on the training set. During training, we randomly crop 10-second segments; utterances shorter than 10 seconds are padded with 
silence. At inference, arbitrary-length audio is processed without padding or chunking.

\subsection{Hyperparameters}
\label{app:hp}
Table~\ref{tab:codec_train_hparams} and Table~\ref{tab:lm_train_hparams} summarize the hyperparameters used for EntangleCodec and the downstream audio language models.

\begin{table}[!t]
  \centering
  \footnotesize
  \setlength{\tabcolsep}{4pt}
  \renewcommand{\arraystretch}{1.02}
  \begin{tabularx}{\columnwidth}{@{}p{0.52\columnwidth}X@{}}
    \toprule
    \textbf{Hyperparameter} & \textbf{Value} \\
    \midrule
    $D_{enc}$             & 768 \\
    $D_{align}$           & 512 \\
    $D_{quant}$           & 14 \\
    GPUs                  & 8 NVIDIA A100 80GB \\
    Optimizer             & AdamW \\
    AdamW $\beta_1, \beta_2$ & 0.9, 0.999 \\
    Weight decay          & $10^{-2}$ \\
    Gradient clipping     & 1.0 \\
    Stage 1 learning rate & $2 \times 10^{-4}$ \\
    LR scheduler          & ReduceLROnPlateau \\
    Scheduler patience    & 5 epochs \\
    LR reduction factor   & 0.5 \\
    Stage 2 learning rate & $1 \times 10^{-4}$ \\
    Per-GPU batch size    & 32 \\
    Gradient accumulation & 4 \\
    Global batch size  & 1024 \\
    Stage 1 steps         & 500k \\
    Stage 2 steps         & 200k \\
    Conditional dropout   & 10\% \\
    \bottomrule
  \end{tabularx}
   \caption{Training hyperparameters of EntangleCodec.}
  \label{tab:codec_train_hparams}
\end{table}

\begin{table}[!t]
  \centering
  \footnotesize
  \setlength{\tabcolsep}{2.5pt}
  \renewcommand{\arraystretch}{0.92}
  \begin{tabularx}{\columnwidth}{@{}p{0.28\columnwidth}p{0.36\columnwidth}X@{}}
    \toprule
    \textbf{Component} & \textbf{Hyperparameter} & \textbf{Value} \\
    \midrule
    \multirow{5}{0.28\columnwidth}{Audio LM Understanding}
      & Base model            & Qwen3-0.6B \\
      & Pretraining steps     & 50k \\
      & Pretraining LR        & $3 \times 10^{-5}$ \\
      & SFT steps             & 10k \\
      & SFT LR                & $1 \times 10^{-5}$ \\
    \midrule
    \multirow{5}{0.28\columnwidth}{Audio LM Generation}
      & Base model            & Qwen3-0.6B \\
      & Tasks                 & TTS and TTA \\
      & Training epochs       & 3 \\
      & Batch size            & 256 \\
      & Learning rate         & $3 \times 10^{-5}$ \\
    \midrule
    \multirow{2}{0.28\columnwidth}{Inference}
      & CFG scale $\gamma$    & 1.0 \\
      & Vocoder               & Vocos \\
    \bottomrule
  \end{tabularx}
   \caption{Training and inference hyperparameters of downstream audio language models.}
  \label{tab:lm_train_hparams}
\end{table}

\subsection{Adversarial Decoder Refinement}
\label{app:gan}

In Stage~2, we introduce an adversarial objective to improve the perceptual quality of reconstructed Mel-spectrograms. The unified encoder, vector quantizer, and a pre-trained discriminator are kept frozen, and only the flow-matching decoder is updated. The discriminator acts as a fixed perceptual critic, providing adversarial supervision by comparing reconstructed Mel-spectrograms with ground-truth ones. This encourages the decoder to produce reconstructions that are both close to the target and perceptually realistic.

The objective of Stage 2 is
\begin{equation}
    \mathcal{L}_{\mathrm{stage2}}
    =
    \mathcal{L}_{\mathrm{flow}}
    +
    \mathcal{L}_{\mathrm{adv}},
\end{equation}
where all loss weights are set to 1.0. Since the encoder and quantizer remain frozen, the adversarial signal only affects the decoder and does not modify the learned semantic-acoustic token representation.

\section{Dataset Details}
\label{sec:data}

\subsection{Training Corpus Statistics}

Table~\ref{tab:dataset} summarizes the datasets used to train EntangleCodec, covering three audio domains. In total, the training corpus comprises approximately 3,200 hours of audio.
\subsection{Rich Audio Caption Generation}
\label{sec:Caption}

A central contribution of EntangleCodec is the use of rich audio captions rather than ASR transcripts for semantic alignment. We generate captions using MIMO-Audio~\cite{mimo} with a structured prompt designed to elicit descriptions across four semantic dimensions: 

(1)~\textbf{Speaker attributes} --- gender, estimated age, accent, speaking style, emotional state; 

(2)~\textbf{Acoustic environment} --- indoor/outdoor, reverberation level, presence and type of background noise; 

(3)~\textbf{Musical attributes} --- tempo, key, instrumentation, mood, dynamics; 

(4)~\textbf{Sound events} --- event type, approximate onset and offset, spatial characteristics. For music-only clips, speaker attributes are omitted; for speech-only clips, musical attributes are omitted. 

For LibriSpeech and LibriTTS, the original ASR transcript is appended to the caption as an additional semantic anchor.

The following examples illustrate the qualitative difference between ASR transcripts and our rich captions across all three audio domains.

\subsection{LLM Training Data}

Table~\ref{tab:llm_data} summarizes the datasets used for audio language model training across understanding and generation tasks.


\begin{table*}[htb]
  \centering
  \footnotesize
  \setlength{\tabcolsep}{10pt}
  \begin{tabular}{llrrr}
    \toprule
    \textbf{Domain} & \textbf{Dataset} & \textbf{Hours} & \textbf{Clips} & \textbf{Avg.\,Dur.\,(s)} \\
    \midrule
    \multirow{2}{*}{Speech}
      & LibriSpeech~\cite{librispeech} & 960.0  & 281,241 & 12.3 \\
      & LibriTTS~\cite{libritts}       & 585.0  & 147,616 & 14.3 \\
    \midrule
    Music
      & MusicBench~\cite{musicbench}   & 327.0  &  52,768 & 22.3 \\
    \midrule
    \multirow{3}{*}{Sound}
      & AudioSet~\cite{audioset}       & 1,154.0 & 415,440 & 10.0 \\
      & AudioCaps~\cite{audiocaps}     &  105.8  &  38,118 & 10.0 \\
      & WavCaps~\cite{wavcaps}         & 127.0  &  45,720 & 10.0 \\
    \midrule
    \textbf{Total}  & & \textbf{3258.8} & \textbf{980,903} & \\
    \bottomrule
  \end{tabular}
  \caption{Training data statistics for EntangleCodec.}
  \label{tab:dataset}
\end{table*}

\begin{table}[!t]
  \centering
  \footnotesize
  \setlength{\tabcolsep}{3pt}
  \renewcommand{\arraystretch}{1.08}
  \begin{tabularx}{\columnwidth}{@{}p{0.18\columnwidth}X r@{}}
    \toprule
    \textbf{Task} & \textbf{Dataset / Type} & \textbf{Samples} \\
    \midrule
    \multirow{3}{0.18\columnwidth}{\makecell[l]{Under-\\standing}}
      & MusicQA~\cite{mullama} / Music QA & 70,011 \\
      & ClothoAQA~\cite{clothoaqa} / Sound QA & 21,114 \\
      & COMPA-R~\cite{gama} / Audio reasoning & 200,234 \\
    \midrule
    \multirow{2}{0.18\columnwidth}{TTS}
      & LibriSpeech~\cite{librispeech} / (text, speech) & 281,241 \\
      & LibriTTS~\cite{libritts} / (text, speech) & 354,780 \\
    \midrule
    \multirow{2}{0.18\columnwidth}{TTA}
      & IFCaps~\cite{ifcaps} / (caption, audio) & 24,000 \\
      & AudioCaps~\cite{audiocaps} / (caption, audio) & 38,118 \\
    \bottomrule
  \end{tabularx}
  \caption{Audio language model training data.}
  \label{tab:llm_data}
\end{table}

\section{Baselines}
\label{app:baseline}
We establish baselines along three evaluation dimensions: tokenizer reconstruction quality, audio understanding, and audio generation.

\paragraph{Tokenizer Reconstruction Quality.}
We compare EntangleCodec with seven state-of-the-art neural audio codecs: DAC~\cite{dac}, EnCodec~\cite{encodec}, WavTokenizer~\cite{wavtokenizer}, SpeechTokenizer~\cite{speechtokenizer}, XCodec~\cite{xcodec}, Mimi~\cite{mimi}, and XCodec2~\cite{xcodec2}. All codec baselines are evaluated using their official pretrained checkpoints under matched compression settings whenever applicable.

\paragraph{Audio Understanding.}
We evaluate EntangleCodec under a codec-controlled setting by training identical Qwen3-0.6B~\cite{qwen3} language models on discrete tokens produced by different codecs, while keeping the model architecture, training data, and optimization setup fixed. This isolates the effect of tokenizer quality on downstream understanding performance. In addition, we compare with specialized continuous-representation audio LLMs equipped with dedicated audio encoders, including LTU~\cite{ltu}, LTU-AS~\cite{ltu-as}, Audio Flamingo~\cite{audioflamingo}, MU-LLaMA~\cite{mullama}, GAMA~\cite{gama}, SALMONN~\cite{SALMONN}, and Qwen2-Audio~\cite{qwen2audio}, using official checkpoints or reported results when available.

\paragraph{Audio Generation.}
For generation tasks, we follow the same codec-controlled protocol as in audio understanding: identical LLM architectures are trained on discrete tokens produced by different codecs, with the training and decoding setup kept fixed. This allows us to directly assess how tokenizer quality affects downstream generation performance.

\section{Benchmark Details}
\label{sec:bench}

\subsection{Audio Understanding Benchmarks}

\paragraph{MMAR.}
MMAR~\cite{mmar} is a multi-domain audio reasoning benchmark containing 1,000 single-choice questions across three domains: speech (covering speaker characteristics, emotion, prosody, and dialogue understanding), music (covering music theory, instrument recognition, and mood), and environmental sound (covering event recognition, scene classification, and causal reasoning). Each question has four answer choices. We report top-1 accuracy following the standard zero-shot evaluation protocol.

\paragraph{MMAU-mini and MMAU.}
MMAU~\cite{mmau} is a large-scale audio understanding benchmark with 9,000 questions in the test split spanning sound events, music, and speech. MMAU-mini is a stratified subset of 1,000 questions designed for rapid evaluation. All models are evaluated zero-shot with top-1 accuracy as the primary metric.

\paragraph{Example Questions.}
Figures~\ref{fig:mmar-example-1},~\ref{fig:mmar-example-2} and~\ref{fig:mmar-example-3} illustrate the type of reasoning required by each benchmark and contrast the responses of EntangleCodec-LLM with a strong baseline.

\begin{figure*}[ht]
\centering
\begin{tcolorbox}[
  colback=blue!3!white, colframe=blue!35!black,
  title={\textbf{Speech Caption Example} --- LibriSpeech, male speaker, read speech},
  fonttitle=\small\bfseries, fontupper=\small,
  left=5pt, right=5pt, top=4pt, bottom=4pt
]
\textbf{ASR Transcript:}\\
``He looked out across the grey expanse of water and said nothing for a long time.''\\[5pt]
\textbf{Rich Caption (Ours):}\\
``A middle-aged male speaker with a deep, measured voice reads \textbf{``He looked out across the grey expanse of water and said nothing for a long time"} aloud at a slow,
deliberate pace. The tone is calm and somewhat solemn, conveying a reflective mood.
The speaker has a mild British Received Pronunciation accent. The recording is clean
and studio-quality, with very low background noise and minimal room reverberation.
No music or non-speech sound events are present.''
\end{tcolorbox}
\caption{Example of a rich speech caption generated by our captioning pipeline.}
\label{fig:speech-caption-example}
\end{figure*}

\begin{figure*}[ht]
\centering
\begin{tcolorbox}[
  breakable,
  colback=green!3!white, colframe=green!40!black,
  title={\textbf{Music Caption Example} --- MusicBench, piano solo},
  fonttitle=\small\bfseries, fontupper=\small,
  left=5pt, right=5pt, top=4pt, bottom=4pt
]
\textbf{Rich Caption (Ours):}\\
``A solo piano piece in a minor key, played at a slow to moderate tempo of approximately
72 BPM. The melody is introspective and melancholic, with a sparse left-hand accompaniment
using sustained bass notes. Dynamic range is moderate, with soft passages punctuated by
occasional crescendos. The recording has a natural concert-hall reverb. No vocals or
other instruments are present. The overall mood is contemplative and slightly sorrowful.''
\end{tcolorbox}
\caption{Example of a rich music caption generated by our captioning pipeline.}
\label{fig:music-caption-example}
\end{figure*}

\begin{figure*}[ht]
\centering
\begin{tcolorbox}[
  breakable,
  colback=orange!3!white, colframe=orange!50!black,
  title={\textbf{Sound Caption Example} --- AudioSet, urban environment},
  fonttitle=\small\bfseries, fontupper=\small,
  left=5pt, right=5pt, top=4pt, bottom=4pt
]
\textbf{Rich Caption (Ours):}\\
``A busy urban street soundscape recorded outdoors. Continuous traffic noise forms
the background, with cars passing at irregular intervals. A car horn honks twice
in quick succession approximately 2 seconds into the clip. Faint footsteps on
pavement are audible in the right channel. The acoustic environment has moderate
urban reverberation, consistent with a narrow street between medium-height buildings.
No speech or music is present.''
\end{tcolorbox}
\caption{Example of a rich sound caption generated by our captioning pipeline.}
\label{fig:sound-caption-example}
\end{figure*}

\begin{figure*}[ht]
\centering
\begin{qualpromptbox}{Example 1 --- Speech: Behavior Recognition}
\promptrole{Audio}
A clip of a male speaker performing a vocal imitation, with exaggerated
tonal patterns and rhythmic speech, recorded in an indoor environment.
\tcbline
\promptrole{Question}
Why did the speaker sing a song at the end?\\
\textbf{(A)}~To express awe for the lion \\
\textbf{(B)}~To showcase his musical talent \\
\textbf{(C)}~To imitate Africans
\tcbline
\promptrole{Ground Truth:} (C) To imitate Africans\\[2pt]
\promptrole{EntangleCodec-LLM (0.6B):} \textbf{(C) To imitate Africans} \textcolor{deltagreen}{$\checkmark$}\\[2pt]
\promptrole{XCodec2-LLM (0.6B):} (B) To showcase his musical talent 
\textcolor{deltared}{$\times$}\\[2pt]
\promptrole{Speechtokenizer-LLM (0.6B):} (B) To express awe for the lion \textcolor{deltared}{$\times$}
\tcbline
\textit{The rhythmic and tonal characteristics of the vocal performance carry
cultural imitation cues. EntangleCodec's rich caption supervision enables
the model to distinguish intentional imitation from musical performance.}
\end{qualpromptbox}
\caption{MMAR Example 1 --- Speech: Behavior Recognition.}
\label{fig:mmar-example-1}
\end{figure*}

\begin{figure*}[ht]
\centering
\begin{qualpromptbox}{Example 2 --- Music: Humor Reduction}
\promptrole{Audio}
A clip of a musical piece with a comedic tone, featuring a notable
pitch pattern that contributes to its humorous effect.
\tcbline
\promptrole{Question}
How to make this song less funny?\\
\textbf{(A)}~Change the note F\# to F \\
\textbf{(B)}~Change the key from F major to C major \\
\textbf{(C)}~Change the note D to C\# \\
\textbf{(D)}~Change the note F\# to D
\tcbline
\promptrole{Ground Truth:} (D) Change the note F\# to D\\[2pt]
\promptrole{EntangleCodec-LLM (0.6B):} \textbf{(D) Change the note F\# to D} \textcolor{deltagreen}{$\checkmark$}\\[2pt]
\promptrole{XCodec2-LLM (0.6B):} (A) Change the note F\# to F \textcolor{deltared}{$\times$}
\tcbline
\textit{The humorous effect stems from the unexpected use of F\#, which creates
a comedic tension. Replacing it with D resolves this tension naturally.
EntangleCodec's fine-grained pitch supervision enables the model to identify
the specific note responsible for the comedic character of the piece.}
\end{qualpromptbox}
\caption{MMAR Example 2 --- Music: Humor Reduction.}
\label{fig:mmar-example-2}
\end{figure*}

\begin{figure*}[ht]
\centering
\begin{qualpromptbox}{Example 3 --- Sound: Fine-grained Audio Discrimination}
\promptrole{Audio}
A clip ending with a brief water-related sound, requiring fine-grained
discrimination between two acoustically similar sound events.
\tcbline
\promptrole{Question}
Is the sound at the end of the audio pouring water or spitting water?\\
\textbf{(A)}~Pouring water sound \\
\textbf{(B)}~Spitting water sound
\tcbline
\promptrole{Ground Truth:} (B) Spitting water sound\\[2pt]
\promptrole{EntangleCodec-LLM (0.6B):} \textbf{(B) Spitting water sound} \textcolor{deltagreen}{$\checkmark$}\\[2pt]
\promptrole{XCodec2-LLM (0.6B):} (A) Pouring water sound \textcolor{deltared}{$\times$}
\tcbline
\textit{The two sounds share similar acoustic properties but differ in subtle
spectral and temporal patterns. EntangleCodec's fine-grained audio caption
supervision enables the model to capture these nuanced distinctions,
while XCodec2-LLM conflates the two acoustically similar sound events.}
\end{qualpromptbox}
\caption{MMAR Example --- Sound: Fine-grained Audio Discrimination.}
\label{fig:mmar-example-3}
\end{figure*}

\subsection{Generation Benchmarks}

\paragraph{SEED-TTS.}
SEED-TTS~\cite{seedtts} evaluates TTS systems on a diverse set of 3,000 utterances spanning multiple speakers and speaking styles. We report WER measured by Whisper-large-v3~\cite{whisper} as the primary intelligibility metric, and UTMOS for naturalness.

\paragraph{AudioCaps (TTA).}
The AudioCaps~\cite{audiocaps} test set contains 979 clips, each paired with five human-written captions. We use the first caption as the generation condition. CLAP Score is computed as the cosine similarity between the CLAP text embedding of the condition and the CLAP audio embedding of the generated output.

\paragraph{Clotho (TTA).}
The Clotho~\cite{clotho} evaluation split contains 1,045 clips with five captions each. We follow the same evaluation protocol as AudioCaps. Clotho covers a broader range of acoustic scenes and environmental sounds with longer, more descriptive captions, providing a complementary evaluation to AudioCaps.

\paragraph{LibriTTS Reconstruction.}
To complement the LibriSpeech reconstruction results in Table~\ref{tab:reconstruction}, we additionally evaluate all tokenizers on LibriTTS \textit{test-clean}~\cite{libritts}. LibriTTS features longer utterances (average 14.3\,s vs.\ 12.3\,s for LibriSpeech) and more diverse speaking styles including expressive and conversational speech, providing a more challenging reconstruction testbed. Results are reported in Table~\ref{tab:libritts}.


The LibriTTS results are consistent with those reported on LibriSpeech in the main paper: EntangleCodec achieves an UTMOS of \textbf{3.94}, within 0.06 points of the top-performing Xcodec2 (4.00) and substantially ahead of reconstruction-focused codecs such as DAC (1.28) and EnCodec (1.54). The slight overall drop compared to LibriSpeech results (e.g., WavTokenizer: 3.79 $\to$ 3.76) is consistent across all methods, attributable to the greater prosodic variability and longer utterance lengths in LibriTTS.


\begin{table}[!t]
  \centering
  \footnotesize
  \setlength{\tabcolsep}{4pt}
  \begin{tabular}{lcccc}
    \toprule
    \textbf{Model} & \textbf{UTMOS}$\uparrow$ & \textbf{F1}$\uparrow$ 
                   & \textbf{STOI}$\uparrow$ & \textbf{SIM}$\uparrow$ \\
    \midrule
    GT              & 4.06 & 0.98 & 1.00 & 1.00 \\
    \midrule
    DAC             & 1.28 & 0.97 & 0.61 & 0.24 \\
    EnCodec         & 1.54 & 0.92 & 0.75 & 0.23 \\
    WavTokenizer    & 3.76 & 0.98 & \textbf{0.89} & 0.63 \\
    SpeechTokenizer & 1.25 & 0.97 & 0.63 & 0.16 \\
    Xcodec          & 3.40 & 0.97 & 0.84 & 0.46 \\
    Mimi            & 3.09 & 0.99 & 0.85 & 0.50 \\
    Xcodec2         & \textbf{4.00} & \underline{0.99} & \underline{0.87} & \textbf{0.75} \\
    \midrule
    \rowcolor{ours}
    \textbf{EntangleCodec} & \underline{3.94} & \textbf{0.99} & 0.86 & \underline{0.65} \\
    \bottomrule
  \end{tabular}
  \caption{Reconstruction quality on LibriTTS \textit{test-clean}. 
           \textbf{Bold} = best; \underline{underline} = second best (GT excluded).}
  \label{tab:libritts}
\end{table}

\section{Evaluation Metrics}
\label{app:metrics}

\paragraph{Reconstruction Quality.}
We evaluate tokenizer reconstruction using complementary objective and perceptual metrics. \textbf{UTMOS}~\cite{utmos} is a deep learning-based predictor of mean opinion score (MOS) for speech quality, ranging from 1 to 5, where higher is better. \textbf{STOI} measures the intelligibility of speech signals, ranging from 0 to 1, where higher values indicate better intelligibility. \textbf{F1 Score} evaluates voiced/unvoiced frame classification accuracy, where higher is better. For sound and music, we additionally report \textbf{AudioBoxScore}~\cite{audiobox}, which averages four sub-scores to assess overall audio quality across domains.

\paragraph{Audio Understanding.}
We report \textbf{Accuracy} for audio understanding tasks, following the standard evaluation protocol of each benchmark.

\paragraph{Audio Generation.}
For TTS, we report \textbf{WER} (Word Error Rate), measured by an automatic speech recognition system, where lower is better. We also report UTMOS for naturalness assessment. For TTA, we report \textbf{CLAP Score}~\cite{clap}, which measures semantic alignment between generated audio and the conditioning text description.


\section{Token Space UMAP Visualization}
\label{sec:viz}

\begin{figure*}[t]
    \centering
    \begin{subfigure}[t]{0.49\textwidth}
        \centering
        \includegraphics[width=\linewidth]{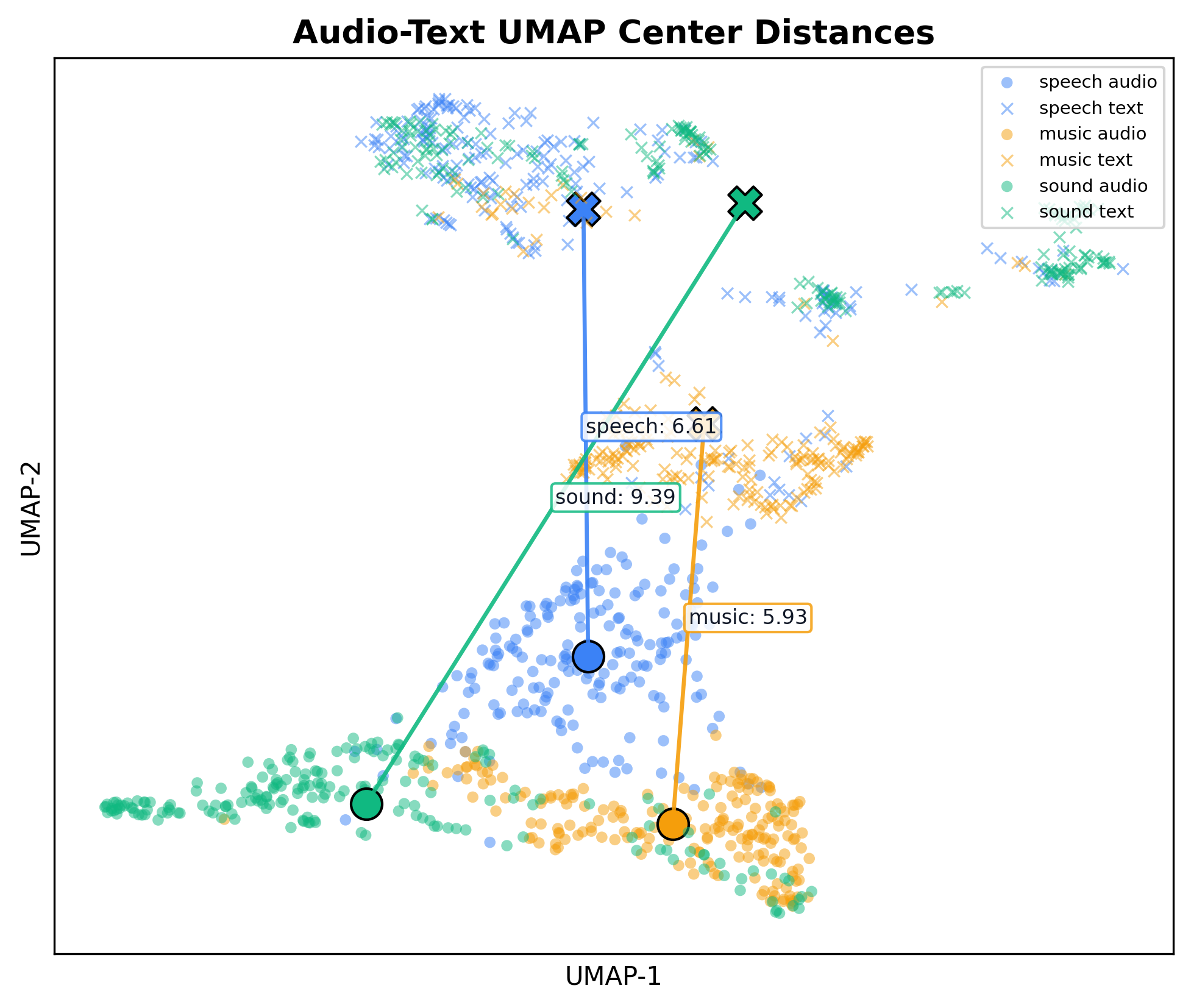}
        \caption{Audio--text UMAP center distances.}
        \label{fig:umap1}
    \end{subfigure}
    \hfill
    \begin{subfigure}[t]{0.49\textwidth}
        \centering
        \includegraphics[width=\linewidth]{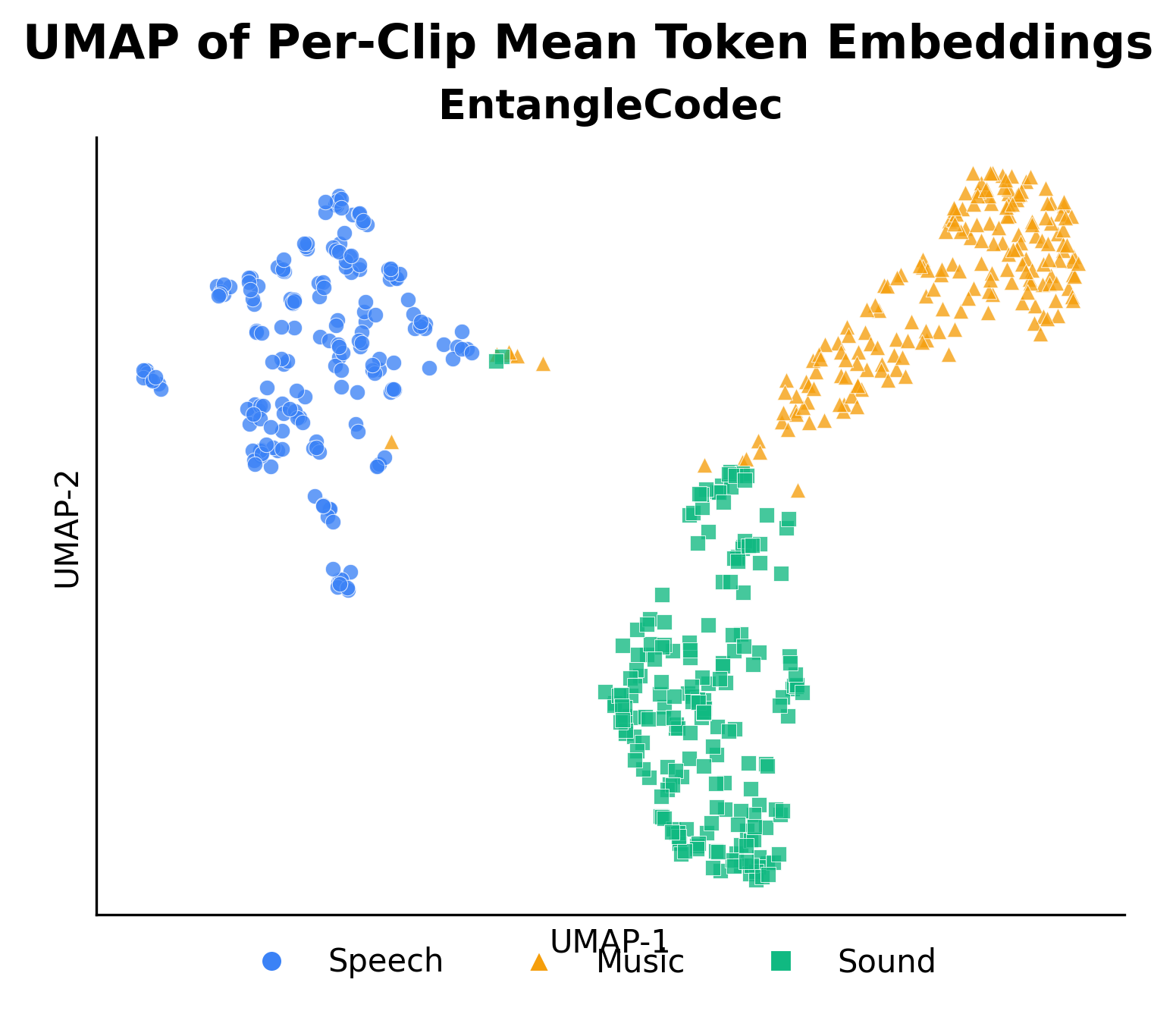}
        \caption{Per-clip mean token embeddings.}
        \label{fig:umap2}
    \end{subfigure}
    \caption{
    UMAP visualization of EntangleCodec representations.
    Left: audio and caption embeddings are projected into a shared space, where smaller center distances indicate closer audio--text alignment.
    Right: per-clip mean token embeddings form clear domain-level clusters for speech, music, and general sound, with additional intra-domain structure.
    }
    \label{fig:umap}
\end{figure*}

We visualize EntangleCodec representations with UMAP to examine whether the learned tokens capture meaningful semantic structure. For each audio clip, we compute the mean token embedding and project it to two dimensions. As shown in Figure~\ref{fig:umap}, EntangleCodec forms clear domain-level clusters for speech, music, and general sound, while also exhibiting meaningful intra-domain organization, such as acoustic/electronic music and slow tempo patterns.

We further visualize the joint audio--text embedding space used for caption alignment. The audio and caption clusters are closely aligned across domains, indicating that rich caption supervision encourages the tokenizer to encode semantic information beyond reconstruction-oriented acoustic detail. These visualizations support the quantitative results, showing that EntangleCodec learns discrete representations that are both acoustically grounded and semantically structured.

\end{document}